\documentstyle[12pt,epsf]{article}
\newcommand{\be}{\begin{equation}}
\newcommand{\ee}{\end{equation}}
\newcommand{\bea}{\begin{eqnarray}}
\newcommand{\eea}{\end{eqnarray}}
\newcommand{\vs}[1]{\vspace{#1 mm}}
\newcommand{\hs}[1]{\hspace{#1 mm}}
\newcommand{\ba}{\begin{array}}
\newcommand{\ea}{\end{array}}
\newcommand{\beann}{\begin{eqnarray*}}
\newcommand{\eeann}{\end{eqnarray*}}
\newcommand{\ct}{\cite}
\newcommand{\r}{\ref}
\newcommand{\nn}{\nonumber \\}
\newcommand{\n}{\nonumber}
\newcommand{\la}{\label}
\renewcommand{\a}{\alpha}

\newcommand{\e}{\epsilon}

\newcommand{\diag}{{\rm diag.}}

\newcommand{\one}{{\mathbf 1}}

\newcommand{\rot}[3]{\left[{#1}\atop{#2}\right]_{#3}}

\newcommand{\cn}{{\cal N}}

\newcommand{\bZ}{{{\bf Z}}}

\makeatletter
\begin{document}

\topmargin 0pt

\oddsidemargin -3.5mm

\headheight 0pt

\topskip 0mm \addtolength{\baselineskip}{0.20\baselineskip}
\begin{flushright}
APCTP-99-008 \\
SOGANG-HEP 259/99 \\
OU-HET 318 \\
{\tt hep-th/9904181}
\end{flushright}
\vs{5}
\begin{center}
{\large \bf Maxwell Chern-Simons Solitons from Type IIB String Theory}\\
\vs{5}
Bum-Hoon Lee~\footnote{bhl@ccs.sogang.ac.kr}\\ 
{\it Department of Physics, Sogang University, Seoul 121-742, Korea}\\
\vs{5}
Hyuk-jae Lee~\footnote{lhjae@phya.yonsei.ac.kr}\\
{\it Department of Physics and Institute of Natural Science,
Yonsei University, \\
Seoul 151-742, Korea} \\
\vs{5}
Nobuyoshi Ohta~\footnote{ohta@phys.sci.osaka-u.ac.jp}\\
{\it Department of Physics, Osaka University, Toyonaka,
Osaka 560-0043, Japan}\\
\vs{5}
Hyun Seok Yang~\footnote{hsyang@physics3.sogang.ac.kr}\\
{\it Department of Physics and Basic Science Research Institute,
Sogang University,\\
C.P.O. Box 1142, Seoul 100-611, Korea}\\
\vs{5}
{\bf ABSTRACT}
\end{center}
We study various three-dimensional supersymmetric Maxwell Chern-Simons
solitons by using type IIB brane configurations. We give a systematic
classification of soliton spectra such as topological BPS vortices and
nontopological vortices in $\cn=2,3$ supersymmetric Maxwell Chern-Simons
system via the branes of type IIB string theory.
We identify the brane configurations with the soliton spectra of the field
theory and obtain a nice agreement with field theory aspects.
We also discuss possible brane constructions for BPS domain wall solutions.
\vs{5}
\begin{flushleft}
April, 1999 \\
\end{flushleft}

\newpage
\section{Introduction}

The recent developments in non-perturbative string theories have
provided new powerful tools to understand the supersymmetric gauge
theories~\ct{GK}. The low-energy dynamics of the D-branes is described by
the supersymmetric gauge theories which can be related to the ground-state
excitations of fundamental strings connecting pairs of D-branes~\ct{witten96}.
The BPS brane configurations in the background led to many exact results
on the vacuum structure of the supersymmetric gauge theories.

Novel aspects of three-dimensional supersymmetric gauge theories
can be understood via type IIB brane configurations, in which D3-branes
are suspended between two NS5-branes~\ct{HW,oz}. This construction gives
an explanation of mirror symmetry in three dimensions via $SL(2,\bZ)$
duality of type IIB string theory. This mirror symmetry is also true for
BPS vortices and exchanges particles and vortices~\ct{ohiss,ks}.

Recently three-dimensional gauge theories have been studied and
classified by using more general type IIB brane configurations, in which
D3-branes are suspended between an NS5-brane and a $(p,q)$5-brane~\ct{ohta}.
In these brane configurations, the three-dimensional field theories, in
general, turned out to be supersymmetric Maxwell Chern-Simons gauge theories
with $\cn=4, 3, 2, 1$ supersymmetry. The $\cn=4$ supersymmetry can
only be realized in NS5-D3-NS5 configuration, which gives supersymmetric QED
without Chern-Simons term. The NS5-D3-$(p,q)$5 configuration gives
$\cn=3, 2, 1$ supersymmetric Maxwell Chern-Simons theory, where
the $\cn=3$ case~\ct{kl,hk} is not much known although it is interesting.

The $\cn=2$ supersymmetric Maxwell Chern-Simons theory is comparably
well known and their soliton solutions have been considerably
studied over the years~\ct{n=2,llw,ivan,leemin}.
Whereas the Maxwell-Higgs model supports only electrically neutral vortices
as topologically stable soliton solutions~\ct{nov}, the addition of
the Chern-Simons term gives rise to topologically stable solutions that
are electrically charged and carry magnetic flux and non-zero angular
momentum~\ct{csv,jlw}. In this theory, there exist topological as well
as nontopological BPS multisoliton solutions since, in three dimensions,
the superpotential allows symmetry broken and unbroken vacua~\ct{llw}.
Thus, there can be a peculiar solution known as the (topological) domain wall
which is a one-dimensional object in three dimensions~\ct{jlw,kll}.
In crossing the domain wall, the vacua are different on two sides.
In addition, it is known that there can also be (nontopological) domain walls
residing in the symmetric phases~\ct{jlw,kll}.
In this paper we will discuss how these kinds of topological objects can be
described in terms of the above brane configurations.

The organization of the paper is as follows.
In sect.~2, by using the similar method taken in refs.~\ct{ohta,ot},
we classify BPS configurations consisting of relatively rotated
two M5-branes with $N_c$ M2-branes in between and $N_f$ M5-branes as well as
other M2-branes corresponding to solitons in three dimensions.
We identify possible supersymmetry for each M-brane configuration.
These are then transformed to the brane configurations in type IIB string
theory after compactifying the M-theory and then applying T-duality.
This construction will provide the classification of all possible BPS solitons
such as topological and nontopological BPS vortices in three-dimensional
supersymmetric Maxwell Chern-Simons system. Although there is a BPS M2-brane
which may be a plausible candidate for the BPS domain wall constructed in
field theory~\ct{jlw,kll}, we have not been able to identify an appropriate
configuration for finite energy density solution.
In sect.~3, we study topological BPS vortices as well as nontopological
vortices by using the M-brane configurations constructed in sect.~2.
We compare the brane configurations with the soliton spectra already known
from field theories and obtain a nice agreement with field theory results.
In sect.~4, we summarize our main results and give qualitative arguments
on the topological as well as nontopological BPS domain wall solutions.
We also discuss mirror symmetry for the solitons,
and indicate future directions.

After the completion of this paper, we were informed from K. Ohta~\ct{ko}
that he also independently arrived at similar results on the moduli
space of vacua in theories considered here.

\section{BPS brane configurations}

The authors of ref.~\cite{ohta} examined three-dimensional gauge dynamics by
using type IIB brane configurations. They obtained these from the M-theory
configurations of M2-branes suspended between two M5-branes at angles~\ct{ot}.
The BPS brane configurations in supersymmetric M-brane backgrounds can
be obtained by the following intersection rules~\ct{ir}:
In the M2-brane background, an M2-brane probe can preserve $1/4$
supersymmetry only without overlap and an M5-brane probe can only in string
intersection. In the M5-brane background, an M5-brane probe can preserve $1/4$
supersymmetry only in string or three brane intersections and an M2-brane
probe can only in string intersection. These situations give
various brane configurations and residual supersymmetries.

In this section, we construct the brane configurations corresponding
to three-dimensional gauge theories with soliton solutions. For this purpose
we need to count the number of supersymmetries remaining in the brane
configurations. The cases of our interest are realized by inserting
other M2-branes intersecting the ${\widetilde {\rm M5}}$-branes that give
rise to the hypermultiplets. After compactification on $x^{\natural}$
(the symbol $\natural$ indicates the eleventh direction, 10) and T$_2$-duality
(the subscript 2 stands for the direction of T-duality) of the brane
configurations, these are reduced to the type IIB brane configurations.
The M2-branes are transformed to a D1-brane or a D3-brane which correspond
to the vortex or domain wall solutions, respectively.
We now explain each case separately.

First let us consider an M2-brane and ${\widetilde {\rm M5}}$-branes between
two M5-branes with relative angles and another ${\widetilde {\rm M2}}$-brane
in the directions $x^2$ and $x^{a}$ (the superscript $a$ indicates one
direction out of 7, 8, and 9 according to the intersection rules), which
corresponds to the D1(0$a$) string embedded in D5(012789)-brane in type IIB
string theory. The worldvolumes of these branes are given by
\bea
\label{m52}
{\rm M5} &:& (012345), \nn
{\rm M2} &:& (01|6|), \nn
{\rm M5}'&:& \left(01\rot{2}{\natural}{\theta}\rot{3}{7}{\psi}
\rot{4}{8}{\varphi}\rot{5}{9}{\rho}\right),\\
{\widetilde{\rm M5}} &:& (01789\natural),\nn
{\widetilde{\rm M2}} &:& (02a),\nonumber
\eea
where the vertical line in M2-brane worldvolume denotes that 6th direction
is bounded by the two M5-branes, and the vertical arrays in the second
M5$'$-brane worldvolume indicate that the brane is rotated along the planes
by the indicated angles. The ${\widetilde{\rm M2}}$ is necessary for our
purpose but was not in ref.~\cite{ohta}. These branes impose the following
constraints on the Killing spinor $\e$~\cite{ot}:
\bea
\label{c1}
{\rm M5} &:& \Gamma_{012345}\e = \e, \\
\label{c2}
{\rm M2} &:& \Gamma_{016}\e = \e,\\
\label{c3}
{\rm M5}'&:& R\Gamma_{012345}R^{-1}\e = \e,\\
\label{c4}
\widetilde{\rm M2} &:& \Gamma_{02a}\e = \e,\\
\label{c5}
\widetilde{\rm M5} &:& \Gamma_{01789\natural}\e = \e,
\eea
where the rotation matrix for the second M5$'$-brane is parameterized by
the four angles as follows:
\be
\la{rotation}
R=\exp\left\{
\frac{\theta}{2}\Gamma_{2\natural} + \frac{\psi}{2}\Gamma_{37}
+ \frac{\varphi}{2}\Gamma_{48}
+ \frac{\rho}{2}\Gamma_{59} \right\}.
\ee

Since $\Gamma_{012\cdots 9\natural}=1$ and so
$\Gamma_{01789\natural}=\Gamma_{016}\Gamma_{012345}$, the condition~(\ref{c5})
is a redundant one. So we must solve just Eqs.~(\ref{c1})-(\ref{c4})
simultaneously as functions of the four angles $\theta, \psi, \varphi$ and
$\rho$. Since $R\Gamma_{012345}R^{-1}=R^2\Gamma_{012345}$, Eq.~(\ref{c3})
becomes
\bea
(R^2-1)\e = 0.
\label{c3'}
\eea
By a straightforward calculation, we obtain
\bea
R^2-1 &=& 2R\Gamma_{2\natural}\left\{
\sin\frac{\theta}{2}\cos\frac{\psi}{2}\cos\frac{\varphi}{2}\cos\frac{\rho}{2}
-\Gamma_{2\natural 37}
\cos\frac{\theta}{2}\sin\frac{\psi}{2}\cos\frac{\varphi}{2}\cos\frac{\rho}{2}
\right. \nn
&&-\Gamma_{2\natural 48}
\cos\frac{\theta}{2}\cos\frac{\psi}{2}\sin\frac{\varphi}{2}\cos\frac{\rho}{2}
-\Gamma_{2\natural 59}
\cos\frac{\theta}{2}\cos\frac{\psi}{2}\cos\frac{\varphi}{2}\sin\frac{\rho}{2}
\nn
&&+\Gamma_{3748}
\sin\frac{\theta}{2}\sin\frac{\psi}{2}\sin\frac{\varphi}{2}\cos\frac{\rho}{2}
+\Gamma_{4859}
\sin\frac{\theta}{2}\cos\frac{\psi}{2}\sin\frac{\varphi}{2}\sin\frac{\rho}{2}
\nn
&& \left.+\Gamma_{3759}
\sin\frac{\theta}{2}\sin\frac{\psi}{2}\cos\frac{\varphi}{2}\sin\frac{\rho}{2}
-\Gamma_{2\natural 374859}
\cos\frac{\theta}{2}\sin\frac{\psi}{2}\sin\frac{\varphi}{2}\sin\frac{\rho}{2}
\right\}.
\label{r2}
\eea

The gamma matrices appearing in the spinor constraints commute with each
other except $\Gamma_{02a}$. Since the square of the matrices is unity and
the traces of their products vanish, we can arrange these matrices by the
same method as in refs.~\ct{ohta,ot} in the following forms:
\bea
\Gamma_{012345} &=& \diag\left(\one_{16},-\one_{16}\right), \nn
\Gamma_{2\natural 37}&=&\diag\left(\one_{8},-\one_{8}, \cdots \right), \nn
\Gamma_{2\natural 48}&=&\diag\left(\one_{4},-\one_{4},\one_{4},-\one_{4},
 \cdots \right), \nn
\Gamma_{2\natural 59}&=&\diag\left(\one_{2},-\one_{2},\one_{2},
-\one_{2},\one_{2},-\one_{2},\one_{2},
-\one_{2}, \cdots \right),
\label{ga2}
\eea
where $\one_{n}$ denotes $n \times n$ identity matrix, and the rests of
Eq.~(\ref{r2}) are determined by the products of the above matrices.
Since $\Gamma_{012\cdots 9\natural}=1$, $\Gamma_{016}$ is also determined
by the products of the gamma matrices in Eq.~(\ref{ga2}) as
\be
\Gamma_{016}-1 =
 -2\times\diag({\bf 0}_2,\one_{2},\one_{2},{\bf 0}_2,\one_{2},{\bf 0}_2,
{\bf 0}_2,\one_{2},\cdots).
\label{e1}
\ee

On the gamma matrix basis~(\r{ga2}), we have the following expression:
\bea
\la{r2ep}
R^2-1 = 2R\Gamma_{2\natural} \hs{-3}
&\times& \hs{-3} \diag\left(
\sin\left(\frac{\theta-\psi-\varphi-\rho}{2}\right)\one_{2},
\sin\left(\frac{\theta-\psi-\varphi+\rho}{2}\right)\one_{2},
\right. \nn
&&\hs{10}
\sin\left(\frac{\theta-\psi+\varphi-\rho}{2}\right)\one_{2},
\sin\left(\frac{\theta-\psi+\varphi+\rho}{2}\right)\one_{2},\nn
&&\hs{10}
\sin\left(\frac{\theta+\psi-\varphi-\rho}{2}\right)\one_{2},
\sin\left(\frac{\theta+\psi-\varphi+\rho}{2}\right)\one_{2},\nn
&&\hs{10}
\left.
\sin\left(\frac{\theta+\psi+\varphi-\rho}{2}\right)\one_{2},
\sin\left(\frac{\theta+\psi+\varphi+\rho}{2}\right)\one_{2}, \cdots
\right).
\eea
Considering the above expression of $\Gamma_{016}$, the remaining
supersymmetry is now determined by the $\sin$ functions of the four angles
in the 1st, 4th, 6th and 7th blocks of $R^2-1$ matrix.
We summarize in Table 1 the BPS brane configurations at angles and
the various supersymmetric theories in three dimensions
obtained in ref.~\cite{ohta} in the absence of the ${\widetilde {\rm M2}}$,
where we indicated only one representative in each case since (3,7)-, (4,8)-,
and (5,9)-planes are on an equal footing with each other.
{\small
\begin{table}
\begin{tabular}{|c|c|c|c|c|l|}
\hline
& angles & condition & SUSY & d=3 & M5$'$\\
\hline
\hline
1 &$\theta(2\natural)$ & $\theta=0$ & $\frac{1}{4}$ & $\cn=4$
 & NS5$\left(12345\right)$ \\
\hline
2-(i) & $\varphi(48), \rho(59)$ & $\rho=\varphi$ & $\frac{1}{8}$ &
$\cn=2$
 & NS5$\left(123\rot{4}{8}{\varphi}\rot{5}{9}{\varphi}\right)$ \\
\hline
2-(ii) & $\theta(2\natural),\rho(59)$ & $\rho=\theta$ & $\frac{1}{8}$
& $\cn=2$
 & $(p,q)5\left(1234\rot{5}{9}{\theta}\right)$ \\
\hline
3-(i) & $\psi(37), \varphi(48), \rho(59)$ & $\rho=\psi+\varphi$
 & $\frac{1}{16}$ & $\cn=1$ & NS5$\left(12\rot{3}{7}{\psi}\rot{4}{8}{\varphi}
 \rot{5}{9}{\psi+\varphi}\right)$ \\
\hline
3-(ii)&  $\theta(2\natural), \varphi(48), \rho(59)$ & $\rho=\theta+\varphi$
 & $\frac{1}{16}$ & $\cn=1$ & $(p,q)5\left(123\rot{4}{8}{\varphi}
 \rot{5}{9}{\theta+\varphi}\right)$ \\
\hline
4-(i)& & $\theta=\psi-\varphi-\rho$ & $\frac{1}{16}$ & $\cn=1$ & $(p,q)5
\left(12\rot{3}{7}{\psi}\rot{4}{8}{\varphi}
\rot{5}{9}{\psi-\varphi-\theta}\right)$ \\
\cline{1-1}\cline{3-6}
4-(ii) & {\small $ \theta(2\natural), \psi(37), \varphi(48), \rho(59)$}
 & $\theta=-\rho, \psi=\varphi$ & $\frac{1}{8}$ & $\cn=2$
 & $(p,q)5\left(12\rot{3}{7}{\varphi}\rot{4}{8}{\varphi}
 \rot{5}{9}{-\theta}\right)$ \\
\cline{1-1}\cline{3-6}
4-(iii) & & $\theta=\psi=\varphi=-\rho$ & $\frac{3}{16}$ & $\cn=3$
 & $(p,q)5\left(12\rot{3}{7}{\theta}\rot{4}{8}{\theta}
 \rot{5}{9}{-\theta}\right)$\\
\hline
\end{tabular}
\caption{Brane configurations at angles and
various supersymmetric theories in 3 dimensions.}
\end{table}
}

Now we consider the BPS configurations constructed by M2-branes such as
${\widetilde {\rm M2}}$ in the M-brane background~(\r{m52}).
We require that (\r{c4}) should not completely break
supersymmetry. Note that the simultaneous solution to Eqs.~(\r{c4}) and
(\r{c3'}) can induce a new constraint depending on their commutativity.
One solution is obtained when the gamma matrices appearing in Eq.~(\r{r2})
all commute with $\Gamma_{02a}$ and we have no further constraint.
The other is when the gamma matrices in Eq.~(\r{r2}) do not commute with
$\Gamma_{02a}$, for which we have an additional condition on the spinor $\e$:
\be
\la{n=3comm}
[R^2, \Gamma_{02a}]\e=0.
\ee
Of course, in this case, the gamma matrices $\Gamma_{02a}$ cannot
be simultaneously diagonalized.

Let us consider the first case. From the expression~(\r{r2}), we see that
we must put at least two angles to zero, resulting in two-angle cases.
There are six possibilities for the gamma matrices to commute with
$\Gamma_{02a}$. By examining each case separately, the $a$ direction is
uniquely determined. The result is the following:
\bea && \theta=\psi=0\; (a=7),\;\;\; \theta=\varphi=0 \;(a=8),
 \;\;\;\theta=\rho=0 \;(a=9),\nn
 && \psi=\varphi=0\; (a=9),\;\;\; \psi=\rho=0 \;(a=8),
 \;\;\;\varphi=\rho=0\; (a=7).
\label{2ang}
\eea

As an example, let us consider the $\theta=\psi=0 \;(a=7)$ case.
In order to find the solution, it is convenient to choose
maximally diagonalized basis different from (\r{ga2}) and (\r{e1}):
\bea
\Gamma_{012345} &=&
\diag\left(\one_{16},-\one_{16}\right), \nn
\Gamma_{4859}&=&\diag\left(\one_{8},-\one_{8}, \cdots \right), \nn
\Gamma_{016}&=&\diag\left(\one_{4},-\one_{4},\one_{4},-\one_{4},
 \cdots \right), \nn
\Gamma_{027}&=&\diag\left(\one_{2},-\one_{2},\one_{2},
-\one_{2},\one_{2},-\one_{2},\one_{2},-\one_{2}, \cdots \right).
\label{ga}
\eea
Using Eqs.~(\ref{ga}), we can rewrite Eq.~(\ref{r2}) as
\be
R^2 -1 =2R\Gamma_{48}\diag\left(\sin\frac{\varphi-\rho}{2} \one_{8},
\sin\frac{\varphi+\rho}{2}\one_{8},\cdots \right).
\ee
On this basis, the first condition (\ref{c1}) kills the second 16 components
of the Killing spinor, and we have to examine the conditions
(\ref{c2}) and (\ref{c3'}) for the first 16 components.
For $\varphi=\rho$ the remaining supersymmetry is reduced by
$\frac{1}{2}$ compared with the configuration without the
$\widetilde{\rm M2}$-branes (case 2-(i) in Table 1) and this brane will
correspond to a BPS state (a vortex) in three dimensions.

Another example is to choose the angles as $\psi=\varphi=0 \;(a=9)$.
We can again arrange the matrices as
\bea
\Gamma_{012345} &=& \diag\left(\one_{16},-\one_{16}\right), \nn
\Gamma_{2\natural 59}&=&\diag\left(\one_{8},-\one_{8}, \cdots \right), \nn
\Gamma_{016}&=&\diag\left(\one_{4},-\one_{4},\one_{4},-\one_{4},
 \cdots \right), \nn
\Gamma_{029}&=&\diag\left(\one_{2},-\one_{2},\one_{2},
-\one_{2},\one_{2},-\one_{2},\one_{2},-\one_{2}, \cdots \right),
\label{ga1}
\eea
Using Eqs.~(\ref{ga1}), we can rewrite Eq.~(\ref{r2}) as
\be
R^2 -1
=2R\Gamma_{2\natural}\diag\left(\sin\frac{\theta-\rho}{2} \one_{8},
\sin\frac{\theta+\rho}{2}\one_{8},\cdots \right).
\ee
For $\theta=\rho$ this case also reduces the supersymmetry by $\frac{1}{2}$
compared with the configuration without the $\widetilde{\rm M2}$-brane
(case 2-(ii) in Table 1) and this also gives a BPS state (a vortex) in
three dimensions. For the remaining cases in Eq.~(\ref{2ang}), we also
obtain similar vortex solutions.

Next we consider the second case. Our interest is in the four-angle cases
corresponding to $\cn=2$ or $\cn=3$ supersymmetry (cases 4-(ii) and (iii)
in Table 1) because BPS states are possible for these cases.
For definiteness, let us choose $a=9$ and the rotation matrix $R^2$ as
\be
\la{4r2}
R^2=\exp\left\{ \theta(\Gamma_{2\natural}-\Gamma_{59})
+\psi(\Gamma_{37}+\Gamma_{48}) \right\}.
\ee
By a straightforward calculation, Eq.~(\r{n=3comm}) becomes
\be
[R^2, \Gamma_{029}]\e=\Gamma_{029}(e^{-4\theta\Gamma_{2\natural}}-1)R^2
\frac{1+\Gamma_{2\natural59}}{2}\e=0,
\ee
which reduces to the following equation:
\be
\la{3to1}
\frac{1+\Gamma_{2\natural 59}}{2}\e=0
\ee
if $\theta \neq n\pi/2 \; (n \in \bZ)$.\footnote{In the case of $\theta=\psi$,
M5$'$-brane is parallel to ${\widetilde{\rm M5}}$-brane in the limit
$\theta \rightarrow \pi/2$ corresponding to 
$\kappa=-\frac{1}{g_s}{\rm tan}\,\theta=-p/q \to \infty$. 
When $\kappa \to \infty$, the vector multiplet decouples and
the theory becomes a theory of a free massless hypermultiplet with
$\cn=4$ supersymmetry~\ct{ks}. It turns out that the supersymmetric pure
Chern-Simons system discussed in refs.~\ct{llw,kl} indeed corresponds to another
limit, i.e., $L_6 \rightarrow 0$ with $\kappa$ fixed, where $L_6$ is the
length of D3-brane in the $x^6$-direction. Thus we here consider only the
case of $\theta \neq \pi/2$.} 
In the four-angle case 4-(ii), one can directly check using Eqs.~(\r{ga2})
and (\r{e1}) that, if the constraints (\r{c1}), (\r{c2}), and (\r{c3'}) are
imposed on the spinor $\e$, the number of unbroken supersymmetry is four
($\cn=2$). The condition (\r{3to1}) does not produce further constraints.
Interestingly, however, the $\cn=3$ supersymmetry $(\theta=\psi$, case 4-(iii)
in Table 1) is further broken to $\cn=2$ by the condition~(\r{3to1}).
Since the gamma matrix $\Gamma_{02a}$ squares to unity and is traceless,
its eigenvalues must be $\pm1$ and the multiplicities of these eigenvalues
should be the same. Moreover, since the traces of products with the gamma
matrices in the spinor constraints vanish, the condition~(\r{c4}) further
breaks supersymmetry at least by half. Consequently, the condition~(\r{c4})
maximally preserves the $\cn=1$ supersymmetry and so
${\widetilde {\rm M2}}$-brane will correspond to a BPS state (a vortex)
in $\cn=2$ or $\cn=3$ supersymmetric theory.
This is consistent with the field theory results in \ct{kl,hk,llw,leemin}.

For other cases with the rotation matrix different from (\r{4r2}),
the direction $a$ of ${\widetilde {\rm M2}}$-brane should be
differently chosen. For example, if we take
\be
R^2=\exp\left\{ \theta(\Gamma_{2\natural}-\Gamma_{37})
+\psi(\Gamma_{48}+\Gamma_{59}) \right\},
\ee
then $a=7$ and similar solutions can be obtained.

For the purpose to construct nontopological vortices, let us consider another
${\widehat {\rm M2}}$-brane in the $(b,\natural)$ directions, instead of
the ${\widetilde {\rm M2}}$-brane in Eq.~(\r{m52}), and then rotate
it by the same rotation $R$ in Eq.~(\r{rotation}) ($b$ indicates one direction
out of 3, 4, and 5 according to the intersection rules).
The Killing spinor condition for this brane is given by
\be
\la{ntvm2}
{\rm {\widehat M2}}: R\Gamma_{0b \natural}R^{-1}\e = \e,
\ee
instead of (\r{c4}).
Note that the rotation in the planes containing neither $b$ nor $\natural$
does not affect the ${\widehat {\rm M2}}$-brane.
What we need for our purpose are not F-strings but D-strings
or their composites, so the angle $\theta$ needs to be nonzero.
We now seek for the simultaneous solutions to Eqs.~(\r{c1}), (\r{c2}),
(\r{c3'}), and (\r{ntvm2}). For definiteness, let us take $b=5$.
Then Eq.~(\r{ntvm2}) is cast into
\be
\la{ntv5}
R_{\theta\rho}^2\Gamma_{05\natural}\e=\e,
\ee
and, from the conditions (\r{c3'}) and (\r{ntv5}), we get another constraint
\be
\la{ntvcomm}
(R_{\theta\rho}^4-1)R_{\psi\varphi}^2\Gamma_{05\natural}\e=0,
\ee
where
\be
R_{\theta\rho}^2=\exp (\theta\Gamma_{2\natural}+\rho\Gamma_{59}),
\;\; R_{\psi\varphi}^2=\exp (\psi\Gamma_{37}+\varphi \Gamma_{48}).
\ee
Since the matrices $R_{\theta\rho},\, R_{\psi\varphi}$, and
$\Gamma_{05\natural}$ are nonsingular, the condition~(\r{ntvcomm}) can be
reduced to the following form:
\be
\la{ntvr2}
(R_{\theta\rho}^2-R_{\theta\rho}^{-2})\e=2\Gamma_{2\natural}
(\sin\theta \cos\rho-\cos\theta \sin\rho\Gamma_{2\natural 59})\e=0.
\ee

In $\cn=2$ theory, in which we put rotation angles not involving $\natural$
and $b$ directions to zero and $\theta=-\rho$, the condition~(\r{ntvr2})
is essentially equal to (\r{c3'}) and so a redundant one.
Using Eq.~(\r{c3'}) or (\r{ntvr2}), the spinor
constraint (\r{ntv5}) reduces to the following condition:
\be
\la{ntvortexm2}
{\rm {\widehat M2}}: \Gamma_{05 \natural}\e = \e.
\ee
Of course, for more general case (\r{ntvm2}), there are three possibilities
for the $b$ direction. The results for each case are the following:
\be
\psi=\varphi=0\; (b=5),\;\;\; \psi=\rho=0 \;(b=4),
 \;\;\;\varphi=\rho=0\; (b=3),
\label{nt2ang}
\ee
and the condition (\r{ntvortexm2}) is generalized to
\be
\la{ntvtm2}
{\rm {\widehat M2}}: \Gamma_{0b \natural}\e = \e.
\ee
If we take the direction of the ${\widehat {\rm M2}}$-brane to satisfy the
condition (\r{nt2ang}), all the gamma matrices appearing in the spinor
constraints commute with each other and so can be simultaneously diagonalized.
We thus see that $\Gamma_{0b\natural}$ can be taken in the same form as
$\Gamma_{029}$ in (\r{ga1}) and the condition (\r{ntvr2}) no longer
breaks supersymmetry. Thus the ${\widehat {\rm M2}}$-brane totally preserves
the $\cn=1$ supersymmetry, giving a BPS state.

On the other hand, in $\cn=3$ theory with $\theta=\psi=\varphi=-\rho$,
the condition (\r{ntvr2}) breaks the supersymmetry from $\cn=3$ to $\cn=2$
as in the topological vortices. Since the condition (\r{ntvortexm2})
further breaks the supersymmetry by half, the ${\widehat {\rm M2}}$-brane
totally preserves the $\cn=1$ supersymmetry,
again giving a BPS state in $\cn=3$ theory.
Since the ${\widehat {\rm M2}}$-branes will be interpreted as
the nontopological vortices, the above results are consistent with the fact
that nontopological BPS vortices can exist only for $\kappa \neq 0$, i.e.,
$\theta \neq 0$.

Finally we consider the other possibility of inserting the second
M2$'$-brane along the $x^5$ and $x^9$ directions, which corresponds to
D3-brane in type IIB string theory. (Of course, we can also choose the
extended directions of the M2$'$-brane to be $(x^3,\,x^7)$ or $(x^4,\,x^8)$
instead of $(x^5,\,x^9)$ according to the intersection rules.)
The condition (\ref{c4}) must be replaced with the condition
\be
{\rm M2}': \Gamma_{059}\e =\e.
\label{d1}
\ee
Contrary to the case (\ref{m52}), $\Gamma_{059}$ commutes with the gamma
matrices in the spinor constraints. All the gamma matrices can be
simultaneously diagonalized and arranged as Eqs.~(\ref{ga2}) and (\r{e1}) and
\be
\Gamma_{059}=\diag\left(1,-1,1,-1,1,-1,1,-1,1,-1,1,-1,1,-1,1,-1, \cdots \right).
\label{e2}
\ee
The supersymmetry is further broken by half by the M2$'$-brane and so the
brane may be a BPS state (a domain wall) in three dimensions.

We have exhausted the M2-brane configurations in the presence of M-brane
background~(\r{m52}) without breaking supersymmetry completely, corresponding
to BPS states.\footnote{One may also consider an M-wave in eleven dimensions,
in which case the Killing spinor condition is $\Gamma_{0\natural}\e=\e$.
The M-wave solution may give a D-string in type IIB string theory
because it reduces to a D0-brane in type IIA string theory.
However, one can see that this solution does not preserve the supersymmetry
in the M-brane background~(\r{m52}). In this paper, we have not considered
M5-brane probes in the M-brane background~(\r{m52}). According to the
intersection rules~\ct{ir}, possible M5-brane probes preserving
$1/4$ supersymmetry are M5$(26ab\natural)$ and M5$(26789)$,
where $a, b=7,\,8,\,9\,(a \neq b)$.}
In the next section, we will realize the BPS soliton states in
three-dimensional field theories in terms of these M-configurations.

\section{Maxwell Chern-Simons vortices}

In this section we will analyze the Maxwell Chern-Simons vortices via type
IIB brane configurations. These in turn can be obtained from the M-brane
configurations constructed in sect.~2 after compactifying the M-configurations
along the eleventh direction and then applying T$_2$-duality~\ct{ohta}.
In the process, the number of unbroken supersymmetries is preserved.

{}From Table 1, we see that, if we set one of the four angles $\theta, \psi,
\varphi, \rho$ in 3-(i) or 3-(ii) to zero, supersymmetry is enhanced from
$\cn=1$ to $\cn=2$, while, in 4-(i) and 4-(ii) cases, it is not enhanced,
for example, in the $\theta \rightarrow 0$ limit where the Chern-Simons term
vanishes. The $\cn=3$ case is quite special since, in this case, the
four angles should be equal.

As noted in sect.~2, in zero- and two-angle cases, there is a possibility
to introduce the ${\widetilde {\rm M2}}$-brane preserving half of the
supersymmetry and extended to $(2\,a)$-plane. In type IIB string theory,
this brane is just the D1-brane along the $a$-direction and, in
three-dimensional field theory, this will correspond to a BPS vortex solution
as we will see. In addition, we have shown that the four-angle cases (4-(ii)
and (iii) cases in Table 1) also contain the spectrum of supersymmetric BPS
vortices. As it will be shown, this fact is consistent with the field theory
result~\ct{leemin,jlw,kl,hk} that the $\cn=2$ and $\cn=3$ Maxwell
Chern-Simons theories admit  topological as well as nontopological vortex
solutions.

As shown in ref.~\ct{schwarz}, the tension of $(q_1,q_2)$-string in the type
IIB metric is
\be
\la{p,q}
T_{(q_1,q_2)}=\frac{1}{2\pi l_s^2}\sqrt{(q_1+q_2\chi)^2+\frac{q_2^2}{g_s^2}},
\ee
where $\chi$ is a constant background of the type IIB RR scalar. 
In the $(p,q)5$-brane background, the instanton coupling on the D3-brane
worldvolume induces the Chern-Simons coupling $\kappa=-\chi$ as discussed
in ref.~\ct{ohta}. In this background, the integer charge $q_1$ is shifted
by an arbitrary amount $\chi$ due to an analogue of Witten's effect that the
electric charge of a monopole is shifted when theta-angle $\theta$
is switched on \ct{dyon}. Thus, although $q_1=0$, a D-string
can carry the electric charge $Q_e$ proportional to the magnetic charge
$Q_m$: $Q_e=|\kappa|Q_m$.

\subsection{Maxwell-Higgs vortices}

Hanany and Witten explained the mirror symmetry in three dimensions through
the $SL(2,{\bf Z})$ duality of type IIB superstring~\ct{HW}.
They considered the supersymmetric configuration with $N_c$ D3-branes in
$(1,\,2,\,6)$ directions suspended between two NS5-branes in
$(1,\,2,\,3, \,4,\,5)$ directions with definite values of $x^6$ coordinate.
This configuration gives $\cn=4$ supersymmetric theory in three dimensions.
We can also construct the gauge field theories with matter fields, if we
insert other $N_f$ D5-branes in $(1,\,2,\,7,\,8,\,9)$ directions preserving
$\cn=4$  supersymmetry. These configurations explain the $\cn=4$ $SU(N_c)$
super Yang-Mill theories with $N_f$ hypermultiplets.
This may be generalized by rotating the second NS5$'$-brane by suitable angles.

First let us consider the brane configuration corresponding to the case 1 in
Table 1 where two NS5-branes are parallel to each other, i.e.,
$\theta=\psi=\rho=\varphi=0$. This corresponds to $\cn=4$ supersymmetric
$SU(N_c)$ gauge theory with $N_f$ hypermultiplets. Here we discuss only
the $N_c=1$ and $N_f=1$ case. This configuration is depicted in Fig.~1-(a)
in which one D3-brane in the direction (126) is suspended between two
NS5-branes in (12345) and intersects with a D5-brane in (12789).
The same is drawn in Fig.~1-(b) when it is seen from the $x^6$ direction.
{}From this configuration, we get U(1) gauge theory with a massless flavor
in the fundamental representation with no Fayet-Iliopoulos (FI) terms.
\begin{figure}[tbp]
\centerline{\epsfxsize=10cm \epsfbox{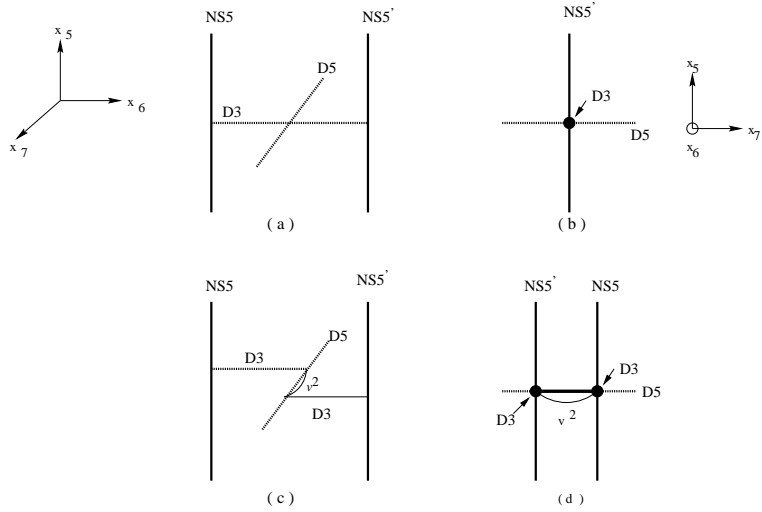}}
\caption{Topological vortices in Maxwell-Higgs theory. (a) or (b) is for
the Coulomb phase and (c) or (d) is a vortex solution in the Higgs phase.}
\label{novortex}
\end{figure}

The $\cn=4$ vector multiplet consists of an $\cn=2$ real vector multiplet $V$
and a chiral multiplet $\Phi$. (For supersymmetric gauge theories,
see, for example, ref.~\ct{wess}.) In $\cn=2$ superspace~\ct{ivan},
the vector multiplet $V$ is composed of $A_\mu \;(\mu=0,1,2)$, which are
the gauge fields on D3-brane worldvolume, and $X_3$ which corresponds to
the $A_3$ component of the four-dimensional gauge field. The chiral multiplet
$\Phi$ contains $X_4$ and $X_5$, which correspond to strings describing
fluctuations of the D3-brane in the transverse directions $(x^4,\, x^5)$.
In addition, there are hypermultiplets consisting of $Q$ and $\tilde Q$ in
the fundamental representation, which originate from the fundamental strings
stretching between the D5- and D3-branes. Using these notations,
we can write down the $\cn=4$ supersymmetric action in the Coulomb branch:
\bea
\label{maxwell}
S_{N{=}4}&=& \frac{1}{g^2} \left[ \int d^3
x d^4 \theta \Phi^{\dagger}\Phi + \frac{1}{2}
   \left( \int d^3 x d^2 \theta W^\a W_\a + {\rm h.c.}\right)
   \right] \nonumber\\
&&+ \int d^3 x d^4 \theta \left(Q^{\dagger}e^{2V}Q
 + \tilde Q e^{-2V} \tilde Q^{\dagger} \right)
+ \frac{1}{\sqrt{2}}\left( \int d^3 x d^2 \theta \tilde Q \Phi Q
 + {\rm h.c.}\right),
\eea
where $W^\alpha$ is the field strength superfield for the real spinor gauge
superfield $U^\alpha(x,\theta)$ and $\a$ is the three-dimensional spinor index.

If we turn on FI couplings for the $\cn=4$ vector multiplets $V$ and $\Phi$
\be
\la{n=4fiterm}
S_{FI}= - v^2\int d^3 x d^4 \theta V - \left( \frac{w^2}{\sqrt{2}}
\int d^3 x d^2 \theta \Phi + \rm{h.c.}\right),
\ee
the scalar potential U in the action (\r{maxwell}) with (\r{n=4fiterm})
is given by
\be
\label{n=4potential}
U=\frac{g^2}{2}\left(|q|^2-|\tilde q|^2 - v^2\right)^2
+\frac{g^2}{2}|q \tilde q - w^2|^2
+\left(|q|^2+|\tilde q|^2\right)\left(X_3^2 +|\phi|^2\right).
\ee
This potential allows only a symmetry broken vacuum:
\be
\la{n=4vacuum}
|q|^2-|\tilde q|^2=v^2,\;\; q\tilde q =w^2,\;\; X_3=\phi=0.
\ee
Note that the FI terms come from the relative
positions in (789)-directions of the NS5- and NS5$'$-branes.
The peculiar fact is that, if the FI coupling $w$ for the complex scalar
field $\Phi$ is nonzero, the hypermultiplet should have all nonzero vacuum
expectation values.

Next we consider the brane configuration corresponding to 2-(i) in Table 1
where the NS5$'$-brane is at angle $\theta=\psi=0$ and $\rho=\varphi$.
This configuration was considered by the authors of refs.~\cite{oz} and
corresponds to $\cn=2$ supersymmetric $U(1)$ gauge theory with a massless
flavor in the fundamental representation with no FI terms.
This configuration is also depicted in Fig.~1-(a) in which one D3-brane
in the direction (126) is suspended between NS5 in (12345) and
NS5$'$-brane in $(123\rot{4}{8}{\varphi}\rot{5}{9}{\varphi})$ and intersects
with a D5-brane in (12789). Note that the masses of matters with flavors
correspond to the position differences in (345)-directions between
the D3- and D5-branes. For this theory, all the terms in Eq.~(\ref{n=4fiterm})
are no longer what would be called FI terms, which break either supersymmetry
or internal symmetry. Because the second NS5$'$-brane is rotated in
(8,9)-directions, the 5-brane shifts in these directions can be compensated
by the D3-brane shifts in the (4,5)-directions, so that it is possible to
preserve both supersymmetry and internal symmetry. On the other hand, the shift
in the relative position in the 7-direction of the NS5- and NS5$'$-branes
corresponds to a real FI term.

In $\cn=1$ superspace, the mass terms for the hypermultiplet are given by
\be
\la{n=2mass}
S_{M}=\int d^3 x d^2 \theta \left(\kappa_4 X_4^2+\kappa_5 X_5^2 \right).
\ee
In Eq.~(\r{n=2mass}), $|\kappa_4|$ and $|\kappa_5|$ correspond to masses for
the scalar fields $X_4$ and $X_5$. These masses originate from the relative
rotations of the NS5$'$-brane in the $(4,\,8)$ and $(5,\,9)$ directions.
The $\cn=2$ supersymmetry requires that the masses of the scalar fields
should be equal to each other, i.e., $\kappa_4=-\kappa_5 \equiv m$.
When this relation is satisfied, Eq.~(\ref{n=2mass}) can be written as
\be
\int d^3x d^2 \theta ( m \Phi^2 + {\rm h.c.}),
\la{ma}
\ee
in $\cn=2$ superspace. We thus see that the shift of the scalar component
of $\Phi$ (corresponding to the D3-brane shifts in (4,5)-directions) cancels
the second terms (linear in $\Phi$) in Eq.~(\ref{n=4fiterm}) and also
produces mass terms for the hypermultiplets from the last term in
Eq.~(\r{maxwell}). As a result, there exists a phase in which the gauge
symmetry is unbroken. This is what we mean when we say that the second terms
in Eq.~(\ref{n=4fiterm}) are not FI terms, and is consistent with our
above brane picture.

Since the only mass scale in this theory is $g^2$, we see, from the
action~(\r{maxwell}) with (\r{ma}), that the mass of the chiral
multiplet $\Phi$ is given by $|m g^2|$ and the hypermultiplets
$Q,\,{\tilde Q}$ and the vector multiplet $V$ are massless. For simplicity,
we take the rotation angle $\varphi=\pi/2$. Since then the mass of the
chiral multiplet $\Phi$ goes to infinity, the chiral field $\Phi$ decouples
from the theory and can be set to zero, but this simplification does not
affect our results.

If we turn on the FI coupling for the vector multiplet by a position
difference between the NS5- and NS5$'$-branes in the 7-direction,
this introduces a linear superpotential in the action (\r{maxwell}) like
\be
\la{fimax}
S_{FI}= - v^2\int d^3 x d^4 \theta V.
\ee
{}From the actions (\r{maxwell}) and (\r{fimax}),
the scalar potential U can be easily read off as
\begin{equation}
\la{maxU}
U=\frac{g^2}{2}\left(|q|^2-|\tilde q|^2-v^2\right)^2+X_3^2(|q|^2+|\tilde q|^2).
\end{equation}
The potential $U$ allows only a symmetry broken vacuum:
\be
\la{sbvacuum}
|q|=v,\;\; |\tilde q|=0,\;\; X_3=0.
\ee
It is well known that the symmetry broken vacuum (\r{sbvacuum})
admits topological Nielsen-Olesen vortices~\ct{weinberg,lmr,leemin},
where it is shown that the mass of $n$ vortices is $2\pi v^2 n$ and
the number of zero modes is $2n$ corresponding to the positions of $n$
vortices. Now we will identify these BPS solutions with the type IIB brane
configurations.

Consider the Higgs branch (\r{sbvacuum}) of the model (\r{maxwell}) sketched
in Fig.~1-(c) and (d). As shown in the figure, the right-hand NS$'$-brane
is shifted by $v^2$ along the 7-direction. This shift introduces the FI D-term
(\r{fimax}). Let us further consider additional D-strings extended to the
7-direction together with the brane configuration in Fig.~1-(c).
Since the D-strings can end on the D3-branes, we can obtain the D-strings with
finite length, which means finite energy. We have shown in sect.~2
that these D-strings preserve half the supersymmetry, and so should correspond
to BPS states. Here we claim that we can identify the D-strings with the
topological Nielsen-Olesen vortices in the Maxwell-Higgs theory. The vorticity
$n$ is just the number of the D-strings. Since $v^2$ has the dimension of
mass and is related to the tension of the stretched D-string, we can interpret
it as the mass of a vortex. From the brane configuration in Fig.~1-(c),
we see that the strings can freely move on the (1,\,2)-plane, so the
translational zero modes of the $n$ D-strings are $2n$.
Thus our identification is consistent with the field theory results.

Note that the chiral multiplet $\Phi$ is neutral with respect to the gauge
group $U(1)$ although it is charged under the $U(1)_{4,5}$ rotation group
in the $(x^4,\,x^5)$ directions. Thus the presence of $\Phi$ in the theory
does not seriously change the story on the existence of the soliton solution.
If we set $\kappa_4=-\kappa_5 \equiv m=0$, the vortex solutions obtained
above can also be considered as the BPS vortices in $\cn=4$ theory. This
class of solutions should be obtained from the BPS solutions in
four-dimensional $\cn=2$ QED.

\subsection{Topological and nontopological Maxwell Chern-Simons vortices}

Here we will analyze the brane configurations 2-(ii) in Table 1 with and
without an additional ${\widetilde {\rm M2}}$-brane extended to
$(2,\,9)$-directions. The corresponding type IIB brane configurations are
depicted in Fig.~2.

Let us identify the three-dimensional $\cn=2$ supersymmetric field theories
realized on the D3-brane. Consider first the configuration in Fig.~2-(a)
in which one D3-brane in the direction (126) is suspended between an NS5 in
(12345) and a $(p,q)5$-brane in $(1234\rot{5}{9}{\theta})$ and intersects
with the $N_f$ D5-branes in (12789). On this configuration, we get U(1)
gauge theory with massless $N_f$ flavors in the fundamental representation
with no FI terms. (Here we will take $N_f=1$ for simplicity.)
Note that the masses of hypermultiplets correspond to
the position differences in (345)-directions between
the D3-brane and the D5-branes and the FI terms come from the relative
positions in (78)-directions of the NS5-brane and the $(p,q)$5-brane. These
FI terms are those with the coefficient $w^2$ in Eq.~(\ref{n=4fiterm}).
In fact, we will see that theories only with the first term have
a symmetry-unbroken phase.
\begin{figure}[t]
\centerline{\epsfxsize=10cm \epsfbox{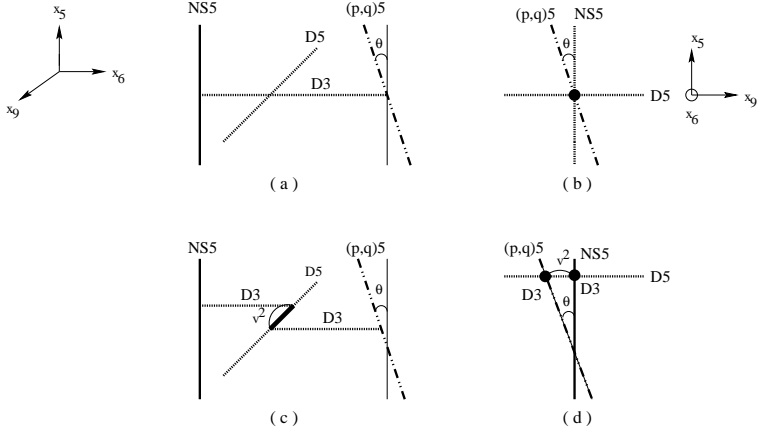}}
\caption{Topological vortices in the Maxwell Chern-Simons theory.
(a) or (b) is for the Coulomb phase and (c) or (d) is a vortex solution in
the Higgs phase.}
\end{figure}

In $\cn=2$ superspace~\ct{ivan}, the vector multiplet $V$ is composed of
$A_\mu$ and $X_5$ corresponding to the $A_3$ component of the four-dimensional
gauge field and the chiral multiplet $\Phi$ contains $X_3$ and $X_4$, which
correspond to strings describing fluctuations of the D3-brane in the
transverse directions $(x^3,\, x^4)$. There are also hypermultiplets
consisting of $Q$ and $\tilde Q$ in the fundamental representation.
Similarly to the Maxwell-Higgs theory in (\r{maxwell}), we can write down
the $\cn=2$ supersymmetric action in the Coulomb branch for the brane
configuration in Fig.~2-(a)~\ct{leemin}:
\bea
\label{n=2mcs}
S_{N{=}2}&=& \frac{1}{g^2} \left[ \int d^3
x d^4 \theta \Phi^{\dagger}\Phi + \frac{1}{2}
   \left( \int d^3 x d^2 \theta W^\a W_\a + {\rm h.c.}\right)
   \right] \nonumber\\
&&+ \int d^3 x d^4 \theta \left(Q^{\dagger}e^{2V}Q
 + \tilde Q e^{-2V} \tilde Q^{\dagger} \right)
+ \frac{1}{\sqrt{2}}\left( \int d^3 x d^2 \theta \tilde Q \Phi Q
 + {\rm h.c.}\right)\\
&&- \frac{1}{2}\int d^3x d^2 \theta \left(\kappa_0
U^{\alpha}W_{\alpha}-\kappa_5 X_5^2 \right)\n.
\eea
In Eq.~(\r{n=2mcs}), $\kappa_0$ and $\kappa_5$ correspond to masses for
the gauge field $A_\mu$ and the scalar field $X_5$. These masses originate
from the simultaneous rotations of the $(p,q)$5-brane in the $(2,\,\natural)$
and (5,\,9) directions. The $\cn=2$ supersymmetry requires that the masses
of the gauge field and the scalar field should be equal to each other, i.e.,
$\kappa_0=\kappa_5 \equiv \kappa$.\footnote{Note that the coupling constant
$\kappa=-\frac{1}{g_s}{\rm tan}\,\theta=-p/q$ is dimensionless, so our $\kappa$
corresponds to $\kappa/g^2$ in~\ct{hk,leemin,jlw}. In ref.~\ct{hk,leemin},
supersymmetric pure Chern-Simons system was obtained by taking the limit
$\kappa \rightarrow \infty$ with the ratio $\kappa/g^2$ fixed. In our
notation, their limit corresponds to $\kappa={\rm fixed}$ and
$g^2\rightarrow \infty$. Since $1/g^2=L_6/g_s$, supersymmetric pure
Chern-Simons theory can be obtained by taking the limit $L_6 \to 0$.
As discussed in ref.~\ct{ohta}, since $x^6$-dependent Kaluza-Klein
modes can be ignored as long as $\kappa \ll 1/g_s$, the low-energy
approximation in the three-dimensional gauge theory should be valid
if the string coupling constant $g_s$ is sufficiently small.}
Since we set the hypermultiplet masses and FI couplings to zero, the only
mass scale in this theory is $g^2$. In fact, from the action (\r{n=2mcs}),
we see that the mass of the vector multiplet $V$ is given by $|\kappa g^2|$
and the hypermultiplets $Q,\,{\tilde Q}$ and the chiral multiplet $\Phi$
are massless.\footnote{Here and in what follows, we omit factors like $1/4\pi$
associated with $\kappa$ for simplicity.} For simplicity, we fix the location
of the D3-brane in the (3,\,4)-plane and put the chiral multiplet $\Phi$ to
zero.

If we shift the positions between the NS5- and $(p,q)$5-branes in
the 9-direction, this introduces a linear term in the action~(\r{n=2mcs})
\be
\la{fiterm}
S_{FI}= - v^2\int d^3 x d^4 \theta V.
\ee
{}From the actions~(\r{n=2mcs}) and (\r{fiterm}),
the scalar potential U can be easily read off as
\be
\label{mcspotential}
U = \frac{g^2}{2}\left(|q|^2-|\tilde q|^2 - v^2 + \kappa X_5\right)^2
+X_5^2(|q|^2+|\tilde q|^2).
\ee
The potential $U$ allows both symmetry broken and unbroken vacua:
\bea
\la{sbphase}
&&{\rm symmetry\; broken\; phase}:|q|=v,\;\; |\tilde q|=0,\;\; X_5=0,\\
\la{sphase}
&&{\rm symmetry\; unbroken\; phase}:|q|=|\tilde q|=0,\;\; X_5
=\frac{v^2}{\kappa}.
\eea
Notice that the linear term~(\ref{fiterm}) allows symmetry unbroken vacuum
and this agrees with the brane picture that the shift of the 5-branes in the
9-direction is compensated by the D3-brane shift in the 5-direction.
It is well known~\ct{leemin,jlw} that the symmetry unbroken phase admits
nontopological BPS multisoliton solutions,
while the symmetry broken phase admits topological BPS multisoliton solutions.
Now our next goal is to find the type IIB brane realizations for these soliton
solutions.

\subsubsection{Topological vortices}

First we go to the Higgs branch (\r{sbphase}) of the model (\r{n=2mcs})
sketched in Fig.~2-(c). As shown in the figure, the right-hand D3-brane is
slidden by $v^2$ along the 9-direction. This introduces the FI
D-term~(\r{fiterm}). Since the transverse fluctuations of the D3-branes
along the (3,\,4)-plane are highly suppressed, the chiral field $\Phi$ decouples
from the theory and can be set to zero. In this Higgs branch,
the theory is mapped to the ${\cal N}=2$ Maxwell Chern-Simons theory
studied in ref.~\ct{leemin}, where the mass of $n$ vortices is $2\pi v^2 n$ and
the number of zero modes is $2n$ corresponding to the positions of $n$ vortices.

Let us consider the D-strings extended to the 9-direction together with the
brane configuration in Fig.~2-(c). We can now apply the similar logic
as that in sect.~3.1. Since the D-strings ending on the D3-branes have
a finite length, the D-strings have a finite energy proportional to their
length. We have shown in sect.~2 that these D-strings preserve half the
supersymmetry, and so correspond to BPS states. Thus we can identify the
D-strings, i.e. $(0,1)$-strings, with the topological vortices
in the Maxwell Chern-Simons theory. Note that, in the presence of the axion
field $\chi$, the tension formula (\r{p,q}) implies that the vortex also
carries electric charge $Q_e$ proportional to magnetic charge $Q_m$, i.e.
$Q_e=\kappa Q_m$, as an analogue of Witten's effect. Field theoretically,
this is just Gauss law constraint~\ct{csv,jlw}.
The vorticity $n$ is just the number of the D-strings. Since $v^2$ has the
dimension of mass and is related to the tension of the stretched D-string,
we can also interpret it as the mass of a vortex. From the brane configuration
in Fig.~2-(c), we see that the strings can freely move on the (1,\,2)-plane,
so the translational zero modes of the $n$ D-strings are $2n$. Thus our
identification is consistent with the field theory results~\ct{jlw,lmr,leemin}.

\subsubsection{Nontopological vortices}
Next we consider the symmetry unbroken phase (\r{sphase}).
We summarize the corresponding brane configuration in Fig.~3-(a).
The D5-brane is lifted up along the 5-direction by $v^2/\kappa$ relative
to the D3-brane, which gives the field $X_5$ a vacuum expectation
value as in (\r{sphase}). Then Eq.~(\r{mcspotential}) shows that it also
induces the mass $m=|v^2/\kappa|$ to the hypermultiplets $Q$ and $\tilde Q$.
\begin{figure}[t]
\centerline{\epsfxsize=10cm \epsfbox{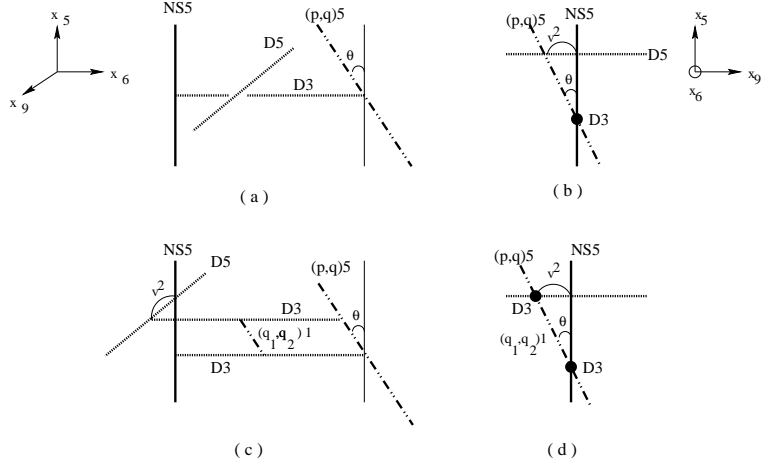}}
\caption{Symmetric phase (a) or (b) and nontopological vortices (c) or (d).}
\end{figure}

Here we will focus on the $b=5$ case out of the ${\widehat {\rm M2}}$-branes
constructed in sect.~2 which may correspond to the nontopological
vortices in theories listed as 2-(ii) in Table 1. Upon compactification,
the ${\widehat {\rm M2}}$-brane reduces to either F-string
in the $\rot{5}{9}{\theta}$-direction or D2-brane in the
$\left(2\rot{5}{9}{\theta}\right)$-directions depending on their worldvolumes,
and further $T_2$-dual transformation gives the F1$\oplus$D1 bound state
in the $\rot{5}{9}{\theta}$-direction. Since the ${\widehat {\rm M2}}$-brane
is rotated from $x^\natural$ by $\theta$ in the $(2,\,\natural)$-plane,
the number $(q_1,q_2)$ characterizing the (F1,D1) bound state satisfies the
relation: $g_s\tan(\frac{\pi}{2}+\theta)=-q_1/q_2$.\footnote{Note the difference
that $(p,q)$ corresponds to (D5,NS5) charges whereas $(q_1,q_2)$ to (F1,D1)
charges, respectively. That this is correct can be understood from the fact
that $\theta=\frac{\pi}{2}$ ($q_1=0$) gives pure D-string in type IIB theory.}
The (F1,D1) bound state has been studied in ref.~\cite{witten96} where
it has been shown that for the configuration of parallel F-string and D-string,
the F-string dissolves in the D-string, leaving flux behind and the resulting
bound state, i.e. D-string with flux, is supersymmetric. In addition, it has
been shown that there is a bound string saturating the BPS bound for all
$(q_1,q_2)$ with relatively prime $q_1$ and $q_2$, named
a $(q_1,q_2)$-string~\ct{schwarz}. In the presence of $(q_1,q_2)$-string,
the axion field in (\r{p,q}) is shifted by $-q_1/q_2$, that is,
$\chi=-\kappa-q_1/q_2$. Thus the nontopological vortex also carries
electric charge $Q_e$ proportional to magnetic charge $Q_m$, i.e.
$Q_e=\kappa Q_m$. All these properties are
consistent with those in field theory results~\ct{csv,jlw}.

Here we propose the type IIB brane configuration for
the nontopological BPS vortices as the form in Fig.~3-(c),
where they are represented by the D-strings with fluxes
connecting two D3-branes.\footnote{The dynamics on upper D3-brane
in Fig.~3-(c) or (d) is not gauge theory but dual scalar theory.
This brane only serves as a boundary state invisible in the lower
D3-brane. On the other hand, the theory on the lower D3-brane is just
our $U(1)$ Maxwell Chern-Simons gauge theory.}
{}From the field theory~\ct{leemin,jlw,lmr}, we know that the magnetic flux
$\Phi$ and electric charge $Q$ of the nontopological vortices are not quantized;
$\Phi=-Q/\kappa=2\pi(n+\alpha)$, where $n$ is the vorticity of the solitons
and $\alpha \ge n+2$ is an undetermined parameter characterizing
the asymptotic behavior of the solutions. It was shown that the number
of zero modes in the nontopological soliton background is $2n+2{\hat \alpha}-2$,
where ${\hat \alpha}$ is the greatest integer less than $\alpha$.
We interpret the number $n$ as the number of D-strings since they can
freely move on the $(1,\,2)$-plane, so the translational zero modes of
the $n$ D-strings are $2n$.\footnote{Although Fig.~3-(c) shows that
the vortices can move along the $x^6$, the moduli of this motion are massive
since $x^6$ has a finite interval.}
In ref.~\ct{jlw}, $2{\hat \alpha}-2$ is interpreted as the moduli parameters
specifying the fluxes and the $U(1)$ phases of lumps. If our identification
is correct, it should be related to the moduli parameters of F-string fluxes.

\subsubsection{$\cn=3$ theories}

Next let us consider BPS vortices in $\cn=3$ theories~\ct{kl,hk}.
The supersymmetry analysis in sect.~2 shows that
the ${\widetilde {\rm M2}}$-brane also corresponds to a BPS vortex solution
preserving $1/16$ supersymmetry in $\cn=3$ theory.
The $\cn=3$ Maxwell Chern-Simons theory was considered in ref.~\ct{hk}
and the action can be obtained from (\r{n=2mcs}) by further adding
mass terms for the chiral multiplets coming from the rotations of
(3,\,7)- and (4,\,8)-planes and the FI couplings of the type (\r{n=4fiterm}).
The scalar potential U for the ${\cal N}=3$ case is given by
\bea
\label{n=3potential}
&&U= \frac{g^2}{2}\left(|q|^2-|\tilde q|^2 - v^2+\kappa X_5\right)^2
+\frac{g^2}{2}|q \tilde q +\kappa \phi- w^2|^2\nn
&&\qquad+\left(|q|^2+|\tilde q|^2\right)\left(X_3^2 +|\phi|^2\right).
\eea
This potential allows a symmetry broken as well as unbroken vacua:
\bea
\la{n=3aphase}
&&{\rm symmetry\; broken\; phase}:|q|^2-|\tilde q|^2=v^2,\;\;
q\tilde q=w^2,\;\;\phi=X_5=0,\\
\la{n=3sphase}
&&{\rm symmetry\; unbroken\; phase}:|q|=|\tilde q|=0,\;\;
\phi=\frac{w^2}{\kappa},\;\; X_5=\frac{v^2}{\kappa}.
\eea

The brane configurations for the BPS vortices in $\cn=3$ theory, for example,
in the case 4-(iii) in Table 1 are essentially the same as those in Figs.~2
and 3. The D-strings in the asymmetric phase (\r{n=3aphase}) and
$(q_1,q_2)$-strings in the symmetric phase (\r{n=3sphase}) correspond
to the topological and nontopological vortices, respectively,
constructed in ref.~\ct{hk} (where the potential $U$ is of the case
$w^2=0$ in (\r{n=3potential})).

As discussed in footnote 7, the vortex solutions of $\cn=2$ and $\cn=3$
supersymmetric Chern-Simons systems considered in refs.~\ct{llw,kl} could
be obtained by taking the limit $L_6 \rightarrow 0$ with $\kappa$ fixed
from the $\cn=2$ and $\cn=3$ Maxwell Chern-Simons theory, respectively.

\section{Discussions}

In this paper we have considered the M-brane configurations which can be
reduced to type IIB branes corresponding to BPS solitons in three-dimensional
gauge theories. In a given M-brane background preserving $\cn=4,\,3,\,2$
supersymmetry, we have found BPS M2-branes preserving $\cn=2,1$ supersymmetry,
where the $\cn=2$ case is obtained only for $\cn=4$ theory and we have
identified the brane configurations with soliton spectra of the field theories.
Although our construction via the type IIB branes can achieve nice agreements
with the vortex solutions of field theory, the type IIB brane construction of
BPS domain wall solutions remain an open problem.
We will briefly discuss BPS M2-branes which are plausible candidates
for the BPS domain wall solutions in field theory.

Type IIB brane construction in this paper will be universally valid
provided that $L_6 < l_s < 1/\kappa g^2$. Thus the supersymmetric Chern-Simons
theories can be obtained by taking the limit $L_6 \rightarrow 0$ from
Maxwell Chern-Simons theories with a Chern-Simons coupling $\kappa$ fixed,
since, in the limit, the mass of gauge boson becomes infinite and so the
kinetic Maxwell term is decoupled. Note, in the limit, that, in the case of
Maxwell theory without Chern-Simons term, the gauge boson remains massless so
that the vector multiplet does not decouple, i.e., the theory flows only to
strong coupling limit. On the other hand, when $\kappa \to \infty$ in
$\cn=3$ theory, the vector multiplet completely decouples and the
supersymmetry is enhanced to $\cn=4$. In the limit, the theory flows to
a free theory of massless hypermultiplets~\ct{ks}.

In three-dimensional Maxwell Chern-Simons theory, it is known that there
can be BPS domain wall solitons~\ct{jlw,kll}: topological domain walls
interpolating the symmetric and asymmetric phases, and nontopological domain
walls residing in the symmetric phase. The domain walls constructed in field
theory are finite energy density solutions.
As also noted in sect.~2, we can introduce the M2$'$-brane
preserving the supersymmetry and extended to $(5,\,9)$-directions.
In type IIB string theory, this brane will be a D3-brane extended in
the $(2,\,5,\,9)$-directions and, in three-dimensional field theory, this
will correspond to a one-dimensional object extended along the $x^2$ direction
(maybe a domain wall). However, there are some problems in the solution.
First of all, the brane configuration do not give finite energy density
solutions. For such solutions, we need D3-branes with finite area in the
$(5,\,9)$-plane. Next, in the cases of ${\cal N}=1$ and ${\cal N}=3$ theory,
there is no explicitly known BPS domain wall solution in field theory whereas
the supersymmetry analysis in sect.~2 shows that the M2$'$-brane preserves
fractional supersymmetry. However, note that the domain wall solution can
be reduced to two dimensional field theory solution as in \ct{kll}.
Then the ${\cal N}=1$ or ${\cal N}=3$ theory corresponds to the two-dimensional
${\cal N}=(1,~1)$ or ${\cal N}=(3,~3)$ supersymmetry, respectively.
Thus, even in these cases, the M2$'$-brane may correspond to the BPS states
in the sense of two-dimensional field theory.

Nevertheless, let us speculate possible brane configurations for the
topological and nontopological domain walls.
Consider the configurations in Figs.~2-(c) and 3-(a)
altogether. The resulting configurations are sketched in Fig.~4.
In sect.~2, we have shown that there can be a BPS state represented by
a D3-brane extended along the (2,\,5,\,9)-directions.
The desired solution is D3-branes confined along the 5 and 9 directions
to obtain a finite energy density solution.
If we could have a D3-brane solution with finite area such as the triangle
in Fig.~4-(b) or (d), its energy density ${\cal E}$, energy per unit length,
will be given by the area of the triangle, i.e., ${\cal E}=v^4/\kappa$,
which is coincident with the field theory result~\ct{jlw,kll}.
\begin{figure}[t]
\centerline{\epsfxsize=10cm \epsfbox{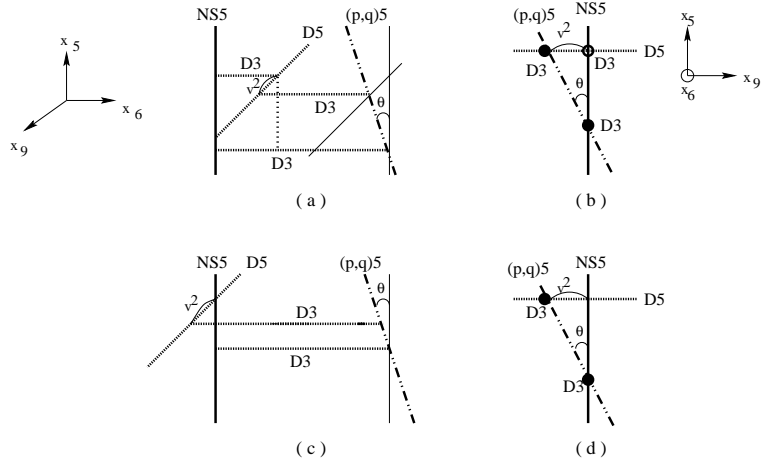}}
\caption{Possible domain walls in Maxwell Chern-Simons theory.
(a) or (b) is topological and (c) or (d) is nontopological.}
\end{figure}

If it is correct, according to the classification in refs.~\ct{jlw,kll},
the solution depicted in Fig.~4-(a) or (b) will correspond to a topological
BPS domain wall since the D3-brane is interpolating the symmetric and
asymmetric phases. On the other hand, the solution in Fig.~4-(c) or (d)
will correspond to a nontopological domain wall since it is residing
only in the symmetric phase.

Unfortunately, it seems that such a D3-brane solution in type IIB supergravity
realizing the BPS domain wall solution with finite energy density has not been
known until now. So at the moment it is
difficult to speculate the problem more precisely.
It will be interesting to look at these problems more closely both
in field theory side and in string theory side.

Finally let us briefly discuss mirror symmetry~\ct{HW,oz,ohiss,ks} for soliton
spectra. Since the mirror symmetry is obtained from the $SL(2,\bZ)$
transformation $S$ of type IIB string theory and a rotation $R$ that maps
$x^i$ to $x^{i+4}\,(i=3,\,4,\,5)$, the combined operation $RS$ exchanges
NS-branes (e.g., NS5-brane and F-string) to
D-branes (e.g., D5-brane and D-string) and maps D3-brane to itself.
Moreover the mirror map exchanges the Higgs and Coulomb branches.
Thus the mirror symmetry transforms a D-string corresponding a soliton
in the Higgs phase into an F-string corresponding to a fundamental particle
in the Coulomb phase. This means that the mirror symmetry exchanges particles
and vortices~\ct{ks,ohiss}. As discussed in refs.~\ct{ohta,ks},
the mirror symmetry transforms Maxwell Chern-Simons theory into self-dual model
with Chern-Simons coupling $\kappa'=-1/\kappa=q/p$ in ref.~\ct{sdm}.
If the mirror symmetry is exact in $\cn=2$ or $\cn=3$ theory, our vortex
construction shows that the nontopological vortices have to transform
into those of the self-dual model represented by $(q_2,-q_1)$-strings.
Thus we expect the mirror map in Maxwell Chern-Simons theory will have
more rich spectrum. It will be interesting to explicitly investigate
the mirror symmetry and the brane creation in refs.~\ct {HW,ohta}
including soliton sectors in the theory.

In this paper we have only considered Abelian gauge theories.
However one may also construct the non-Abelian Yang-Mills Chern-Simons theory
and its Higgs phase via type IIB brane configurations.
(The $\cn=3$ supersymmetric non-Abelian Chern-Simons theory
and its breaking to $\cn=2$ is partially constructed in~\ct{ko}.)
We hope that the generalization of the present work will be achieved
near future.

\newpage
\noindent
{\large\bf Acknowledgments}\\[.2cm]
We would like to thank APCTP for the kind hospitality, where part of this
work was done. B.-H.L. and H.S.Y. were supported in part by BSRI Program
under BSRI 98-2414.
H.-j.L was supported in part by Korea Science and Engineering
Foundation under Grants No. 97-07-02-02-01-3 and the Korea Research
Foundation (1997).
N.O. was supported in part by the Monbusho International
Scientific Research Program: Joint Research
``Fundamental principle beyond the standard model".

\newcommand{\J}[4]{{\sl #1} {\bf #2} (#3) #4}
\newcommand{\andJ}[3]{{\bf #1} (#2) #3}
\newcommand{\AP}{Ann.\ Phys.\ (N.Y.)}
\newcommand{\MPL}{Mod.\ Phys.\ Lett.}
\newcommand{\NP}{Nucl.\ Phys.}
\newcommand{\PL}{Phys.\ Lett.}
\newcommand{\PR}{Phys.\ Rev.}
\newcommand{\PRL}{Phys.\ Rev.\ Lett.}
\newcommand{\PTP}{Prog.\ Theor.\ Phys.}
\newcommand{\ib}{{\it ibid.}}
\newcommand{\hep}[1]{{\tt hep-th/{#1}}}


\end{document}